\documentclass[fleqn,usenatbib]{mnras}

\usepackage{newtxtext}
\usepackage{amsmath}

\usepackage{listings}
\usepackage{xcolor}
\usepackage[capitalize]{cleveref}

\definecolor{codegreen}{rgb}{0,0.6,0}
\definecolor{codegray}{rgb}{0.5,0.5,0.5}
\definecolor{codepurple}{rgb}{0.58,0,0.82}
\definecolor{backcolour}{rgb}{0.95,0.95,0.92}

\lstdefinestyle{mystyle}{
    backgroundcolor=\color{backcolour},   
    commentstyle=\color{codegreen},
    keywordstyle=\color{magenta},
    numberstyle=\tiny\color{codegray},
    stringstyle=\color{codepurple},
    basicstyle=\ttfamily\footnotesize,
    breakatwhitespace=false,         
    breaklines=False,                 
    captionpos=b,                    
    keepspaces=true,                 
    numbersep=5pt,                  
    showspaces=false,                
    showstringspaces=false,
    showtabs=false,                  
    tabsize=2
}

\usepackage[T1]{fontenc}

\usepackage{graphicx}
\usepackage{amsmath}
\usepackage{amssymb}

\title[GW190521 and AGN J124942.3+344929]{Current observations are insufficient to confidently associate the binary black hole merger GW190521 with AGN J124942.3+344929}

\author[Ashton et al. (2020)]{Gregory Ashton,$^{1,2}$\thanks{E-mail: gregory.ashton@ligo.org}
Kendall Ackley,$^{1,2}$
Ignacio Maga\~{n}a Hernandez,$^{3}$
\newauthor
Brandon Piotrzkowski$^{3}$
\\
$^{1}$School of Physics and Astronomy, Monash University, Vic 3800, Australia\\
$^{2}$OzGrav: The ARC Centre of Excellence for Gravitational Wave Discovery, Clayton VIC 3800, Australia\\
$^{3}$University of Wisconsin-Milwaukee, Milwaukee, WI 53201, USA\\
}

\pubyear{2020}

\usepackage{xspace}

\newcommand{\odds}{\mathcal{O}}
\newcommand{\I}{\mathcal{I}}

\newcommand{\gw}{GW190521\xspace}
\newcommand{\ztfname}{ZTF19abanrhr\xspace}
\newcommand{\dGW}{d_{\rm gw}}
\newcommand{\dEM}{d_{\rm em}}

\begin{document}
\label{firstpage}
\pagerange{\pageref{firstpage}--\pageref{lastpage}}
\maketitle

\begin{abstract} 
Recently, \citet{graham2020} identified \ztfname as a candidate electromagnetic counterpart to the binary black hole merger \gw. The authors argue that the observations  are consistent with a kicked binary black hole interacting with the accretion disk of the activate galactic nucleus AGN J124942.3+344929. If a real association (rather than happenstance), this has implications for the sources of LIGO/Virgo binary mergers, future prospects for electromagnetic counterparts, and measurements of the expansion rate of the Universe. In this Letter, we provide an analysis of the multi-messenger coincident-significance based on the localisation overlap and find that that the odds of a common source for \gw and \ztfname range between 1 and $12$ depending on the waveform model used; we consider this insufficient evidence to warrant confidently associating \gw with \ztfname.
\end{abstract}


\section{Introduction}
\gw is a high-mass binary black hole merger observed by the LIGO \citep{aLIGO} and Virgo \citep{aVirgo} gravitational wave detectors \citep{GW190521}. First announced as the public trigger S190521g\footnote{\url{https://gracedb.ligo.org/superevents/S190521g}}, this event is exceptional amongst the events published so far due to its very high mass. The public alert allowed the rapid follow-up of the candidate by optical telescopes; of these, ZTF, the Zwicky transient facility \citep{2019PASP..131a8002B, 2019PASP..131g8001G} reports a candidate counterpart\footnote{\url{https://lasair.roe.ac.uk/object/ZTF19abanrhr/}}, \ztfname. The counterpart is confirmed to be a flare from the active galactic nuclei (AGN) J124942.3$+$344929. The flare, which begins approximately 26~days after the merger of \gw, is argued to be caused by the remnant black hole, kicked through the accretion disk of the AGN \citep{2019ApJ...884L..50M}.

If \gw can be confidently associated with \ztfname, this would be the first association of an electromagnetic counterpart to a binary black hole merger \citep{Perna2019};
although a weak, short electromagnetic transient was observed 0.4s after GW150914 \citep{Connaughton2016}.
This has significant implications: the identification of the AGN, which has a well-measured spectroscopic redshift\footnote{\url{http://skyserver.sdss.org/dr12/en/tools/explore/Summary.aspx?id=1237665128546631763}} of $z=0.438\pm0.00003$ and allows a new standard-candle measurement of the Hubble constant \citep{chen2020, gayathri2020b, mukherjee2020} comparable to that of GW170817 \citep{abbott17_gw170817_Hubble}. In addition, it has implications for the study of the gaseous accretion disk surrounding the AGN and the population properties of the compact binary systems observed by LIGO/Virgo \citep{2019ApJ...884L..50M, 2019PhRvL.123r1101Y}.

In this work, we quantify the probability of a common-source hypothesis [i.e. a binary black hole merger followed by an AGN flare due to the \citet{2019ApJ...884L..50M} mechanism] between \gw and \ztfname based on the source luminosity distance and sky localisation. We do not address whether such a model is physical, we only quantify the agreement based on the observations. Ultimately, the question of whether the observations are due to a common source will best be answered by future observations: \citet{graham2020} make a verifiable prediction of a repeat flare within the next few years. Nevertheless, we hereby aim to make a statement based solely on the initial observations themselves as to whether the two can be confidently associated. We review the Bayesian coincident detection significance method \citep[based on][]{Ashton2018} in \cref{sec:method}, provide results in \cref{sec:results} before concluding in \cref{sec:conclusion}.

\section{Method}
\label{sec:method}
\gw and \ztfname are individually confident detections. But, what is the probability they have a common source compared to the probability that they amount to a random coincidence? Answering this question in general depends on two aspects. First, the physics, how plausible are models which predict an AGN flare due to a kicked binary black hole and what are the rates of those compared with other phenomena which could also explain the data? Second, does the data support the notion that they arise from a common source? To answer the second question, \citet{graham2020} applied a $p$-value approach common in the literature to multi-messenger significance \citep[see, e.g.,][]{abbott17_gw170817_gwgrb}. They estimate the probability that the event is a random coincidence alone. Here, we instead apply a Bayesian approach based on \citet{Ashton2018} which seeks to compare the probability that they do have a common source to the probability that they are a random coincidence (see \citet{budavari2008, naylor2013, budavari2011, budavari2015} for other similar approaches). We will not attempt to address the first question, the physical plausibility of the model, but instead focus only on whether the observations have a common source.

We define two hypotheses. First, a common-source hypothesis, $C$ in which the binary black hole merger and the remnant causes the AGN flare; second, a random coincidence hypothesis, $R$, in which the binary black hole merger and AGN flare are entirely separate events. The ratio of the probabilities comparing these two hypotheses is then given in the usual way \citep[see, e.g.,][]{sivia1996data} by the \emph{odds}:
\begin{align}
    \odds_{C/R} \equiv \frac{P(C | \dGW, \dEM, I)}{P(R | \dGW, \dEM, I)} \pi_{C/R} \,.
\end{align}
The first factor here, is the \emph{Bayes factor}; the ratio of probabilities for the two hypotheses conditional on the two gravitational-wave and electromagnetic data sets $\dGW$ and $\dEM$ and any cogent prior information $I$. The second factor is the prior-odds, $\pi_{C/R}\equiv \pi(C| I) / \pi(R | I)$: the ratio or probabilities for the two hypotheses based solely on the prior information.

The prior odds are subjective and depend on arguments as to the plausibility of the proposed mechanisms by which a binary black hole merger can produce an electromagnetic counterpart. Later, in \cref{sec:results}, we estimate the prior odds using the simulation studies performed by \citet{graham2020}; we will then discuss the sensitivity of our conclusions to this prior choice in \cref{sec:conclusion}.

Binary black hole merger models and the \citet{2019ApJ...884L..50M} AGN flare models share a set of common parameters which can be used to calculate the Bayes factor comparing a common-source with a random coincidence hypothesis. If the two observations arise from from a common source, they should have a common luminosity distance $D_L$ and sky location $\Omega$ (specifically, we define $\Omega$ in coordinates of right ascension and declination). Intuitively then, we want to quantify how well the posterior distributions for these \citep[i.e. Fig.~1 of ][]{graham2020} agree. We use this set of common-model parameters $D_L$ and $\Omega$ to quantify the Bayes factor. 

Within the framework we use \citep[see][]{Ashton2018}, it is simple to include additional common-model parameters (e.g., the time between the merger and flare, or remnant kick velocity). The amount by which they change the overall conclusions is proportional to how well the data constrains the parameters relative to their prior uncertainty. The flare-merger interval and the remnant kick velocity are poorly constrained. This, combined with significant modelling uncertainty leads us to conclude that a simple estimate, based solely on the sky-localisation and distance is more appropriate.

The odds are then calculated from
\begin{align}
    \odds_{C/R} = \pi_{C/R} \I_{D_L ,\Omega} \approx \pi_{C/R} \I_{D_L} \I_{\Omega}\,
    \label{eqn:odds1}
\end{align}
where $\I_{D_L,\Omega}$ is the combined, while $\I_{D_L}$ and $\I_{\Omega}$ are the separate \emph{posterior overlap integrals} \citep{Ashton2018}; for any arbitrary parameter set $\theta$ the posterior overlap integral is given by
\begin{align}
   \I_{\theta} = \int \frac{p(\theta | \dGW, C)p(\theta | \dEM, C)}{\pi(\theta | C)}\,d\theta \,,
\end{align}
where the numerator is the product of the two posterior distributions while $\pi(\theta | C)$ is the common-source hypothesis prior.

The factorisation in \cref{eqn:odds1} into separate sky and distance components effectively discards information from the coincidence calculation about any correlations between $\Omega$ and $D_L$. These parameters are not believed to be strongly correlated, so this is likely a robust approximation, an assumption we validate in \cref{sec:conclusion}.

Since \ztfname is localized significantly better than the gravitational-wave signal (the right ascension, declination of \ztfname and the redshift of the AGN have sub-percentile relative uncertainties), we can treat the posterior distribution condition on $\dEM$ as a delta-function at the transient sky location $\Omega'$ and AGN luminosity distance $D_L'$. Then the odds simplifies to
\begin{align}
    \odds_{C/R} & = \pi_{C/R} \frac{p(D_L', \Omega'| \dGW, C)}{\pi(D_L', \Omega'| C)} = \pi_{C/R} \I_{D_L, \Omega} \\
    & \approx \pi_{C/R} \frac{p(D_L'| \dGW, C)}{\pi(D_L'| C)}\frac{p(\Omega'| \dGW, C)}{\pi(\Omega' | C)} =\pi_{C/R}\I_{D_L}\I_{\Omega}\,.
    \label{eqn:odds}
\end{align}
We calculate the odds using both the combined and separated overlap integral methods in \cref{sec:results} in order to compare results and establish the component that dominates the odds.
The gravitational-wave posterior distributions are provided in \citet{GW190521} as a set of \emph{posterior samples}; we use a Kernel Density Estimate (KDE) method to interpolate these samples when evaluating \cref{eqn:odds}. The program to reproduce our results (both for the combined and separate overlap integrals) is provided in \cref{sec:appendix}.

\section{Results}
\label{sec:results}
In \citet{GW190521}, three \emph{waveform models} are used to analyse the data: NRSur7dq4 \citep{NRsur}, SEOBNRv4PHM \citep{SEOB1, SEOB2} and IMRPhenomPv3HM \citep{IMRPhenomPv3HM}. Using a variety of waveforms allows a study of the systematic uncertainty due to the differing model assumptions. For GW190521, the three waveform gives different estimates of the posterior distribution, though the overall conclusions are broadly consistent \citep{GW190521_implications}. In \cref{tab:results}, we calculate the odds from \cref{eqn:odds} using the posterior distributions for each waveform model and provide useful constituent elements of the calculation. In the following we describe how each constituent is calculated in detail, then discuss the overall conclusion in \cref{sec:conclusion}

\begin{table}
    \centering
    \begin{tabular}{l|r|r|r|r|r}
         Waveform Model&  $\pi_{C/R}$ & $\I_{D_L}$ & $\I_{\Omega}$ & $\I_{D_L}\I_{\Omega}$& $\odds_{C/R}$\\ \hline
         NRSur7dq4 & 1/13 & 1.8 & 29 & 52 & 4.0 \\
         SEOBNRv4PHM & 1/13 & 3.6 & 41 & 150 & 12 \\
         IMRPhenomPv3HM & 1/13 & 1.2 & 22 & 26 & 2.0 \\
    \end{tabular}
    \caption{Odds and constituent elements, see \cref{eqn:odds1} and \cref{eqn:odds}, for the three waveform models used in analysing \gw assuming distance and sky-localization are separable. Values greater than 1 indicate support for the common-source hypothesis $C$, while values less than 1 indicate support for the random-coincidence hypothesis $R$. We estimate the odds are subject to statistical errors from the numerical evidence estimates and reweighting procedure totalling a few percent.}
    \label{tab:results}
\end{table}

The prior-odds quantify the probability of the common-source hypothesis [i.e. a binary black hole and AGN flare consistent with the \citet{2019ApJ...884L..50M} model] compared to a random coincidence hypothesis in which the AGN flare is not related to the binary black hole. In \citet{Ashton2018}, a detailed discussion of the calculation was provided in the context of a short Gamma-ray burst and binary neutron star merger.  It was found that an approximate estimate of the prior odds is given by the inverse of the number of events which could conceivably classified as consistent with the common-source hypothesis in the duration and volume searched. This has the intuitive implication that if $N$ events could be consistent with a common-source hypothesis, the Bayes factor needs to be larger than $1/N$ in order to make a confident association.  Estimating $N$ can be done using information about the rate of such events, but here we can instead use the estimate of the number of flares similar to \ztfname in the ZTF alert stream, which was 13 \citep{graham2020}. Our priors odds can therefore be conservatively (in the sense of favouring the common-source hypothesis) estimated as $\pi_{C/R}=1/13$; we note these prior odds assume the physical plausibility of the model itself.

\begin{figure}
    \centering
    \includegraphics{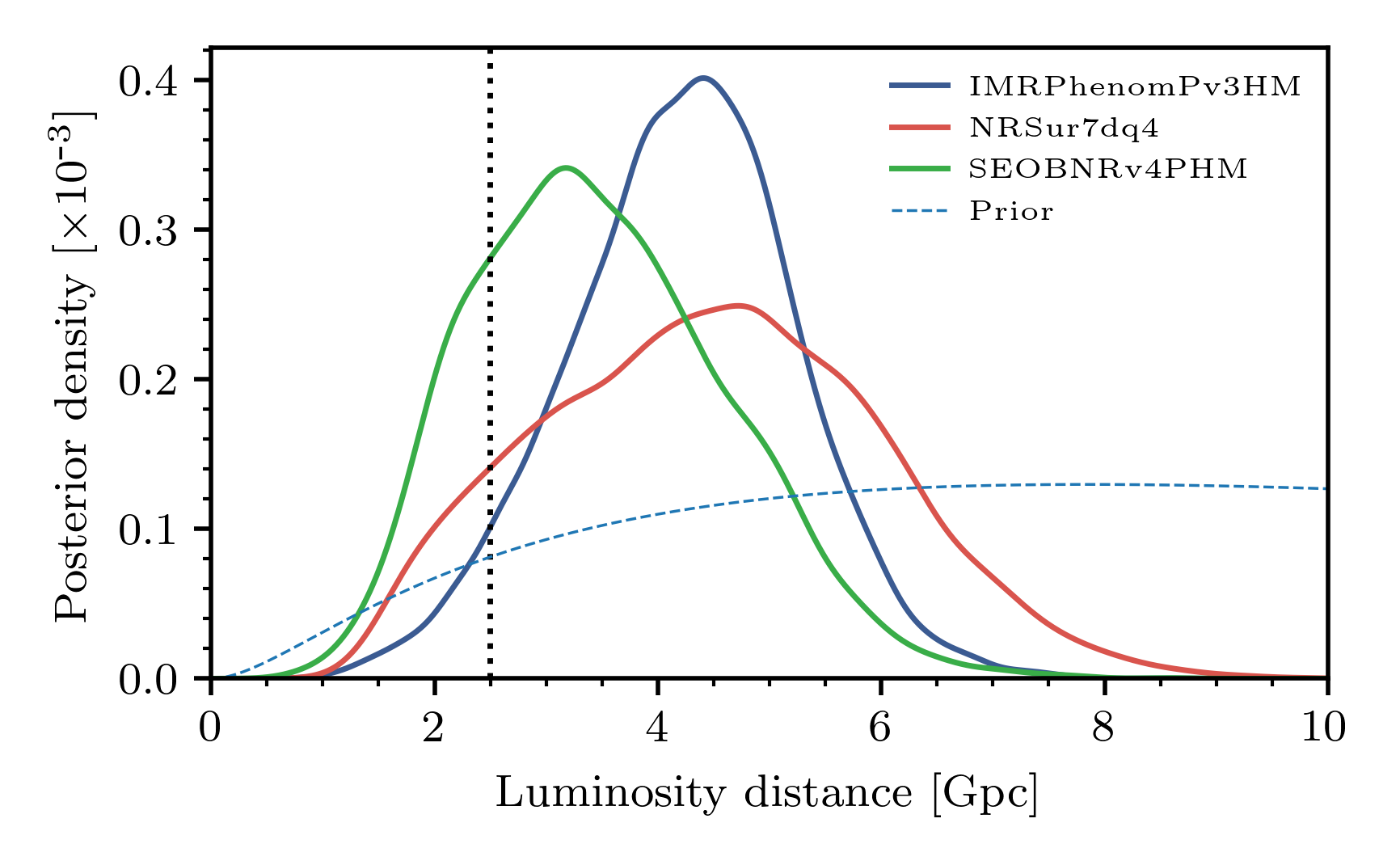}
    \caption{Luminosity distance distribution for the three waveform models used in analysing \gw. A vertical dotted line marks the luminosity distance of AGN J124942.3$+$344929, a dashed line marks the uniform in source-frame prior used in this analysis.}
    \label{fig:redshift}
\end{figure}

The luminosity distance distribution of the three waveforms, along with the position of the AGN are given in \cref{fig:redshift}. The initial analysis of \gw applied a prior uniform in the square of the luminosity distance ($D_L$), i.e. $\pi(D_L|C)\propto D_L^2$.
Beyond a redshift of~1, for which \gw shows some support, cosmological effects become important. To improve the physical plausibility of the common-source prior, we re-weight the \citet{GW190521} posterior distributions to the uniform in source-frame prior described in \citet{bilby2020}, with identical bounds. Throughout the analysis, we use the \citet{plank2015} cosmology. Subsequently, we estimate $\I_{D_L}$ by evaluating a KDE of the posterior distribution at $D_L'$, the location of AGN J124942.3$+$344929. In \cref{tab:results}, across all three waveforms, $\I_{D_L}$ provides fairly weak evidence for the association. Visually, the location of the AGN in \cref{fig:redshift} sits in the bulk of the posterior; $\I_{D_L}$ is the ratio of the posterior to the prior at the same point.

The spatial overlap of \gw and \ztfname can be seen in Fig.~1 of \citet{graham2020} which used the initial skymap produced by LIGO/Virgo. Updated skymaps, based on improved waveform models and better-calibrated data, can be found in \citet{GW190521_implications}, but are broadly consistent with the initial skymap. Using these HEALPix skymaps \citep{HEALPix} and creating one for the AGN using its sky coordinates, we calculated $\I_{\Omega}$ with the method developed for \texttt{RAVEN} \citep{cho2019low}, a low-latency pipeline that searches for gamma-ray or neutrino busts coincidences with LIGO/Virgo/Kagra gravitational wave events. The results of this method are found in \cref{tab:results}. Across all three waveforms, $\I_{\Omega}$ provides moderate evidence for the common-source hypothesis which matches up to visual inspection similar to in \citet{graham2020}. Taking the product of the prior odds and the individual posterior overlap integrals, $\I_{D_{L}}$ and $\I_{\Omega}$, we calculate the overall odds $\odds_{C/R}$ for each of the three waveform models in \cref{tab:results}.

Separating the analysis into contributions from the sky-location and luminosity distance  allows us to understand the contributions from each separately. 
In \cref{tab:results}, the luminosity distance provides weak evidence for the association, with the sky-localisation dominating the combined Bayes factor. The Bayes factor based solely on the distance $I_{D_L}$ is sensitive to the arbitrary upper bound of the prior luminosity distance prior. In our analysis, we use the value 10~Gpc chosen in the original analysis; varying this upper bound within reasonable range of values does not change the overall conclusion. This, along with that the sky-localisation dominates the calculation, verifies that our overall conclusion is robust to this prior sensitivity.

We repeat our analysis without assuming that the luminosity distance and sky-location are independent of each other and separate (see \cref{eqn:odds1}). We approximate the 3-dimensional posterior on luminosity distance and sky-location using the clustered KDE routines available within \texttt{ligo.skymap} \citep{Singer:2016erz}. We then use a 3-dimensional KDE to compute the overlap integral $\I_{D_L,\Omega}$ for each waveform model (see \cref{sec:appendix} for details). We show the results for the odds and the overlap integral in \cref{tab:results_3d}. The significance for the coincidence is reduced by a factor of 1.3-2 depending on the waveform model that is used compared to the results in \cref{tab:results}.

\begin{table}
    \centering
    \begin{tabular}{l|r|r|r}
         Waveform Model&  $\pi_{C/R}$  & $\I_{D_L,\Omega}$ & $\odds_{C/R}$\\ \hline
         NRSur7dq4 & 1/13 & 31 & 2.4 \\
         SEOBNRv4PHM & 1/13 & 120 & 9.2 \\
         IMRPhenomPv3HM & 1/13 & 14 & 1.1 \\
    \end{tabular}
    \caption{Odds and constituent elements for the three waveform models used in analysing \gw without separating distance and sky-location. The value of $\I_{D_L,\Omega}$ is computed using a 3-dimensional clustered KDE in order to fully take into account any correlations between the luminosity distance and sky-location. We estimate the odds are subject to statistical errors from the numerical evidence estimates and reweighting procedure totalling a few percent.}
    \label{tab:results_3d}
\end{table}

The prior odds are necessarily subjective, but the value chosen in this work is conservative in the sense that it avoids penalising too harshly a possible association and therefore favours the common-source hypothesis.
We consider the prior range $\pi_{C/R} < 1$ to be plausible; values larger than one would imply a prior preference for the association which we think to be unreasonable. Our overall conclusion, that the evidence is insufficient to warrant confidently associate the observations, is robust to this range of plausible prior choices.

\section{Discussion and Conclusion}
\label{sec:conclusion}
When combining the prior odds, the contribution from the distance, and the contribution from the sky-location, across all three waveforms, the odds range from $2$ to $12$ (1 to 10 when including correlations between luminosity distance and sky-location). This constitutes evidence \emph{in favour} of associating \gw with the transient \ztfname, but it remains tentative. (For comparison, \citet{Ashton2018} found an odds in excess of $10^{6}$ for the association between GW170817 and GRB~170817A). This also validates our assumption that ignoring correlations between distance and sky-location is reasonable since our overall end result remains the same.

The odds calculated herein are subject to uncertainty which is dominated by the waveform model. The original analysis \citep{GW190521} considered three precessing quasi-circular models, which we find to spread the range of odds by one order of magnitude.
Recent work has shown that the observation is also consistent with non-circular (i.e. eccentric) mergers of black holes \citep{gayathri2020, romero-shaw2020} and the head-on collision of horizonless vector boson stars \citep{CB2020}. Under these models, the measured luminosity distance of the merger is smaller (relative to the precessing quasi-circular waveforms considered in \cref{fig:redshift}), improving the posterior agreement with the luminosity distance of AGN~J124942.3$+$344929. At the moment, it is not possible to measure the luminosity distance while including both the effects of precession and eccentricity because no model is available which combines the two.
However, the luminosity distance of the merger is the subdominant contribution to the overall odds. We estimate that using the luminosity distances measured by eccentric waveform models will change the overall odds by only a factor of a few: insufficient to change our overall conclusions. We estimate other statistical uncertainties due to the sampling and posterior-density approximations to total less than a few percent.

We conclude that the evidence is insufficient to confidently associate the events and hence use the astrophysical implications based on the association.
Nevertheless, the tentative association of \ztfname with \gw represents an exciting development in gravitational-wave astronomy. While we do not find sufficient evidence to confidently associate the events, this should motivate electromagnetic observers to pursue follow-up of future binary black hole events which may shed light on the phenomena. Future observations, with improved sensitivity from detector improvements, may be better localized and hence result in confident association of a high-mass binary black hole with an AGN, hence validating the \citet{2019ApJ...884L..50M} model. On the other hand, the repeat flare predicted by \citet{graham2020} may be observed resulting in a more confident association.

\section{Acknowledgements}
We thank Francesco Pannarale, Colm Talbot, Will Farr, Paul Lasky, and Thomas Dent for useful comments and discussion which improved this work. IMH is supported by the NSF Graduate Research Fellowship Program under grant DGE-17247915. This work makes use of the
\texttt{scipy} \citep{scipy:2020},
\texttt{numpy} \citep{oliphant2006guide, van2011numpy, harris2020array},
\texttt{gwcelery} (\url{https://gwcelery.readthedocs.io/en/latest/}),
\texttt{ligo.skymap} \url{https://lscsoft.docs.ligo.org/ligo.skymap/},
and \texttt{bilby} \citep{bilby} scientific software packages. We use the posterior samples and HEALPix \citep{HEALPix} files for \gw \citet{GW190521} available form \url{https://dcc.ligo.org/LIGO-P2000158/public}.

\bibliographystyle{mnras}
\bibliography{bibliography}

\appendix

\begin{figure*}
\raggedright

\section{Program to evaluate the odds and reproduce the results of Tables 1 and 2}
\label{sec:appendix}

The \texttt{Python3} software packages required to run the program below can be installed with the command
\begin{lstlisting}[language=bash]
$ pip install bilby gwcelery ligo.skymap ligo-raven h5py
\end{lstlisting}
The data required to run this program (\texttt{GW190521\_posterior\_samples.h5}  and \texttt{GW190521\_Implications\_figure\_data.tgz}, which needs to be decompressed) are available from
\url{https://dcc.ligo.org/LIGO-P2000158/public}. The program below assumes the \texttt{h5} and \texttt{fits} files are copied to the directory in which the program is run.
The results obtained herein used results from \href{https://dcc.ligo.org/LIGO-P2000158-v4/public}{LIGO-P2000158-v4}.

\begin{lstlisting}[language=Python, caption={}, label=listing:code]
import numpy as np
from h5py import File
from scipy.stats import gaussian_kde
from bilby.core.prior import Cosine, Uniform, PowerLaw
from bilby.gw.prior import UniformSourceFrame
from bilby.gw.conversion import redshift_to_luminosity_distance
from gwcelery.tasks.external_skymaps import create_external_skymap
from ligo.raven.search import skymap_overlap_integral
from ligo.skymap.io.fits import read_sky_map
from ligo.skymap.kde import ClusteredKDE

np.random.seed(123)  # Fixed seed for reproducibility
# https://lasair.roe.ac.uk/object/ZTF19abanrhr/
ra_em_deg, dec_em_deg = 192.426239, 34.824715
# http://skyserver.sdss.org/dr12/en/tools/explore/Summary.aspx?id=1237665128546631763
z_em = 0.438

# Conversions
ra_em_rad, dec_em_rad = np.deg2rad(ra_em_deg), np.deg2rad(dec_em_deg)
dL_em = redshift_to_luminosity_distance(z_em)
prior_odds = 1 / 13
dfile = File("GW190521_posterior_samples.h5", mode="r")

# Set up prior distributions
LVC_dL_prior = PowerLaw(alpha=2, minimum=1, maximum=10000)
dL_prior = UniformSourceFrame(name='luminosity_distance', minimum=1, maximum=10000)
ra_prior_rad, dec_prior_rad = Uniform(minimum=0, maximum=2 * np.pi), Cosine()

for waveform in ["NRSur7dq4", "SEOBNRv4PHM", "IMRPhenomPv3HM"]:
    data = dfile[waveform]["posterior_samples"]
    ra_samples_rad, dec_samples_rad, dL_samples = data["ra"], data["dec"], data["luminosity_distance"]

    # Re-weight to uniform in source frame prior
    weights = dL_prior.prob(dL_samples) / LVC_dL_prior.prob(dL_samples)
    draws = np.random.uniform(0, max(weights), weights.shape)
    keep = weights > draws
    ra_samples_rad = ra_samples_rad[keep]
    dec_samples_rad = dec_samples_rad[keep]
    dL_samples = dL_samples[keep]

    # Calculate I_Omega
    gw_skymap, header = read_sky_map(f"GW190521_{waveform}_skymap.fits.gz")
    external_skymap = create_external_skymap(ra_em_deg, dec_em_deg, 0., '')
    I_Omega = skymap_overlap_integral(external_skymap, gw_skymap)

    # Calculate I_DL
    I_DL = gaussian_kde(dL_samples)(dL_em)[0] / dL_prior.prob(dL_em)

    # Calculate I_DL_Omega
    pts = np.column_stack((ra_samples_rad, dec_samples_rad, dL_samples))
    theta_em = np.column_stack((ra_em_rad, dec_em_rad, dL_em))
    KDE = ClusteredKDE(pts, max_k=15)
    pi_DL_Omega = np.prod([
        ra_prior_rad.prob(ra_em_rad),
        dec_prior_rad.prob(dec_em_rad),
        dL_prior.prob(dL_em)])
    I_DL_Omega = KDE(theta_em)[0] / pi_DL_Omega
    print(f"{waveform}: IdL={I_DL:0.2g} IOmega={I_Omega:0.2g} IdLOmega={I_DL_Omega:0.2g}")

\end{lstlisting}

\end{figure*}

\bsp
\label{lastpage}

\end{document}